# Mo Atom Rearrangement Drives Layer-Dependent Reactivity in Two-Dimensional MoS$_2$

Zifan Wang[1], Jiaxuan Wen[2], Tina Mihm[3], Shaopeng Feng[2], Kelvin Huang[3], Jing Tang[4], Tianshu Li[4], Liangbo Liang[6], Sahar Sharifzadeh[3], Keji Lai[2], Xi Ling[1, 4, 5*]

[1] Department of Chemistry, Boston University, Boston, Massachusetts 02215, USA.

[2] Department of Physics, The University of Texas at Austin, Austin, Texas 78712, USA.

[3] Department of Electrical and Computer Engineering, Boston University, Boston, Massachusetts 02215, USA.

[4] Division of Materials Science and Engineering, Boston University, Boston, Massachusetts 02215, USA.

[5] The Photonics Center, Boston University, Boston, Massachusetts 02215, USA.

[6] Center for Nanophase Materials Sciences, Oak Ridge National Laboratory, Oak Ridge, Tennessee 37831, USA.

[*]To whom the correspondence should be addressed. Email address: xiling@bu.edu

**Abstract**

Two-dimensional (2D) materials offer a valuable platform for manipulating and studying chemical reactions at atomic level, owing to the ease of controlling their microscopic structure at the nanometer scale. While extensive research has been conducted on the structure-dependent chemical activity of 2D materials, the influence of structural transformation during the reaction remains largely unexplored. In this work, we report the layer-dependent chemical reactivity of $MoS_2$ during a nitridation atomic substitution reaction and attribute it to the rearrangement of Mo atoms. Our results show that the chemical reactivity of $MoS_2$ decreases as the number of layers is reduced in the few-layer regime. In particular, monolayer $MoS_2$ exhibits significantly lower reactivity compared to its few-layer and multilayer counterparts. Atomic-resolution transmission electron microscope (TEM) reveals that MoN nanonetworks form as reaction products from monolayer and bilayer $MoS_2$, with the continuity of the MoN crystals increasing with layer number, consistent with the local conductivity mapping data. The layer-dependent reactivity is attributed to the relative stability of the hypothetically formed MoN phase which retain the number of Mo atomic layers present in the precursor. Specifically, the low chemical reactivity of monolayer $MoS_2$ is attributed to the high energy cost associated with Mo atom diffusion and migration necessary to form multi-layer Mo lattices in the thermodynamically stable MoN phase. This study underscores the critical role of lattice rearrangement in governing chemical reactivity and highlights the potential of 2D materials as versatile platforms for advancing the understanding of materials chemistry at atomic scale.

**Introduction**

Tailoring material structures to manipulate chemical reactions with atomic precision has become a central theme in modern material science.[1–8] Two-dimensional (2D) materials provide a unique platform for studying chemistry at atomic level, due to the ease of control of their microscopic structures such as stacking configuration and thickness. In recent years, growing interest has emerged in exploring the correlation between the microscopic structure and chemical performance of 2D materials such as graphene and $MoS_2$.[9–18] For instance, 1T phase transition metal dichalcogenides (TMDs) exhibit significantly higher reactivity in organohalides surface functionalization reaction, compared to the 2H phase.[19,20] Layer-dependent chemical reactivity of $MoS_2$ was further reported in both ion-intercalation phase transition[21] and the hydrogen evolution reaction (HER)[22], which was attributed to the distinct electronic structure of $MoS_2$ with different numbers of layers. Similarly, strong thickness- and stacking-order-dependent surface functionalization has been observed in graphene.[23,24]

In conventional models of chemical reactivity, the electronic properties of reactants are considered central to determining reaction rates, and this perspective is well captured by Marcus theory.[25–28] Marcus theory has been widely applied to understand the electron-transfer reactions on 2D materials such as carbon nanotubes[29–32] and graphene.[12,24,33,34] However, the theory is based on the premise that the chemical species do not undergo significant lattice rearrangement during electron transfer, limiting its applicability in many reaction systems—particularly in solid-state transformations where the product lattice differs substantially from that of the precursor. For example, in a previously reported atomic substitution reaction converting van der Waals (vdW) layered $MoS_2$ into non-vdW molybdenum nitride (MoN),[35–37] the process involves not only charge transfer between $MoS_2$ and $NH_3$, but also substantial crystal structure transformation from $MoS_2$ to MoN—extending beyond the scope of Marcus theory. Investigating how the reaction kinetics depend on the microscopic structure of $MoS_2$, such as the number of layers, could offer significant insights into the fundamental principles governing material reactivity.

Here, we use the atomic substitution reaction in $MoS_2$ as a model system to demonstrate that its layer-dependent chemical reactivity is driven by the Mo atom rearrangement. We observe that thinner $MoS_2$ (e.g., monolayers and bilayers) exhibit significantly lower chemical reactivity compared to thicker counterparts. Using transmission electron microscope (TEM), we characterize the resulting MoN morphology and find that, while crystalline MoN forms across all precursor thicknesses, the microstructural morphology varies significantly. Specifically, MoN nanonetworks form from the reaction products of monolayer and bilayer $MoS_2$, whereas a continuous MoN film is observed when starting from thicker $MoS_2$. The results are consistent with local conductivity imaging by microwave impedance microscopy (MIM). To elucidate the layer-dependent chemical reactivity and the morphological evolution of the resulting MoN, we propose a mechanistic interpretation centered on the rearrangement of Mo atoms necessary for the formation of thermodynamically stable MoN structures. Density Functional theory (DFT) calculations investigate the initial step of the reaction and the binding mechanism of $NH_3$ to the $MoS_2$ surface.

**Results and Discussion**

We first demonstrate the layer-dependent reactivity on a step-shaped $MoS_2$ flake consisting of monolayer (1L), few-layer (FL) and multilayer (ML) regions. The sample is mechanically exfoliated and transferred onto a $SiO_2$/Si substrate, followed by a nitridation reaction with ammonia gas in a quartz tube furnace. Figure 1a and S1 in the Supplementary Information (SI) show the optical images of the sample at various stages of the reaction. As the reaction proceeds, pronounced changes in optical contrast are observed in the FL and ML regions, indicating the formation of MoN,[35–37] also supported by the X-ray photoelectron spectroscopy (XPS) measurements (Figure S2). In contrast, the monolayer region shows no visible change in optical contrast, suggesting significantly lower reactivity. Notably, the thicker regions exhibit fast and more complete conversion. In the ML region, the reaction initiates within the first 3 mins and is completed by 15 mins. The FL region shows a delayed response, with surface conversion beginning at around 12 mins and completing at 27 mins. Remarkably, the 1L region remains unconverted even after 27 mins, with no observable contrast variation, highlighting its high chemical inertness under the same reaction conditions.

The thickness-dependent reactivity is further characterized by Raman spectroscopy, as shown in Figure 1b to 1d. After 15 mins of reaction, the $E^1_{2g}$ and $A_{1g}$ modes of $MoS_2$ disappear in the ML region, while a new Raman peak at 211 cm$^{-1}$ —attributed to the $Mo_5N_6$ phase[35,38] —emerges, indicating the successful conversion of $MoS_2$. In the FL region, the two Raman modes of $MoS_2$ remain detectable even after 24 mins; however, their intensities significantly decrease, suggesting that a substantial portion of $MoS_2$ has been converted. In addition to the intensity decrease, a red shift of the $E^1_{2g}$ mode is observed, which is attributed to increased strain associated with the formation of shorter Mo-N bonds.[39–41] The conversion in the FL region completes at around 27 mins, as evidenced by the disappearance of $MoS_2$ Raman signals and the appearance of a $Mo_5N_6$ Raman peak at 204 cm$^{-1}$.[38] In contrast, strong and well-defined Raman signals from $MoS_2$ persist in the 1L region after 27 mins of reaction. The slight reduction of intensity and peak broadening are attributed to the degradation of crystallinity and the slow thermal decomposition of monolayer $MoS_2$ under prolonged high temperature annealing.

We further performed TEM characterizations on a partially converted step-shaped $MoS_2$ flake. As shown in Figure 1e, the sample contains three distinct regions with thicknesses of 26.7, 6.3 and 3.3 nm, as measured by atomic force microscopy (AFM) prior to transfer onto a TEM grid. In the 26.7 nm and 6.3 nm regions, MoN-$MoS_2$ heterojunctions are clearly observed, with well-defined interfaces outlined by yellow dashed lines. This is corroborated by the presence of two distinct sets of selected area electron diffraction (SAED) patterns at the junctions. In contrast, no evidence of MoN formation is observed in the thinnest 3.3 nm region, displaying only a single set of SAED patterns corresponding to $MoS_2$. This results further confirm that thinner $MoS_2$ exhibits lower chemical reactivity in the nitridation reaction compared to thicker $MoS_2$.

We further probe the layer-dependent reactivity in the ultrathin regime through a comparative analysis of monolayer (1L), bilayer (2L), trilayer (3L) and quadrilayer (4L) domains. Figure 2a and Figure S4 show optical images of the ultrathin step-shaped $MoS_2$ sample, with each layer-thickness region labeled and separated by white dashed lines. The layer number of each region is determined through Raman and photoluminescence (PL) spectroscopy,[42–44] as shown in Figure S3. Optical microscope, Raman intensity mapping of the $A_{1g}$ mode of MoS$_2$, along with gigahertz conductivity mapping using microwave impedance microscopy (MIM)[45,46] are used to monitor the reaction progress during nitridation. As the reaction proceeds, optical contrast changes—similar to those described in Figure 1a—are observed, further indicating layer-dependent reactivity even within the few-layer regime. In particular, the 3L and 4L regions show the onset of conversion at 6 mins, with significant transformation by 12 mins. In contrast, the 2L and 1L regions convert more slowly, with noticeable contrast changes at 14 and 18 mins, respectively. Although the reactivity difference between 3L and 4L is minimal, pronounced differences are observed among 1L, 2L and 3L regions. Especially, monolayer MoS$_2$ remains highly inert in the conversion, requiring 42 mins for complete conversion, consistent with our earlier findings.

These conclusions can be further supported by the Raman intensity mapping shown in Figure 2b. At 0 min, the Raman intensity increases progressively from the 1L to 4L region, primarily due to the optical interference effect associated with both laser excitation and Raman scattering.[47–49] As the conversion proceeds, a substantial decrease in Raman intensity is observed across all four regions, reflecting the transformation of MoS$_2$. However, the evolution of Raman intensities follows distinct trends in different layers. To elucidate these differences, we extract the average $A_{1g}$ peak intensity from each region and perform a statistical analysis (see Methods), as presented in Figure 2d. At 6 min, the $A_{1g}$ peak intensity decreases in all four regions—not primarily due to chemical conversion, but rather due to crystallinity degradation and the onset of slow thermal decomposition of MoS$_2$, as previously mentioned. The magnitude of intensity reduction exhibits an inverse correlation with layer thickness—1L (25.2%), 2L (21.0%), 3L (18.3%), and 4L (10.9%)—suggesting that thinner MoS$_2$ exhibits lower thermal stability, consistent with prior reports.[18] After an additional 6 min of annealing, a significant Raman intensity drop is observed in the 3L (55.0%) and 4L (49.5%) regions, indicating their conversion to MoN, while the 1L and 2L regions show invisible change, reflecting their lower reactivity. This observation is in excellent agreement with the optical contrast evolution. By 13 min, the $A_{1g}$ peak intensities of both 4L and 3L drop to zero, indicating the full conversion on these two regions. The 2L region shows a significant intensity drop from 81.6% to 49.8% at 14 min, consistent with the partial conversion suggested by the optical images, and reaches full conversion at 18 min, with the intensity decreases to zero. Notably, a residual $A_{1g}$ peak intensity of 18.8% remains in the 1L region at 18 min, indicating incomplete conversion. The Raman intensity of 1L vanishes at 42 min—much later than that for 2L, highlighting remarkably lower reactivity of monolayer $MoS_2$. These results further underscore the pronounced layer-dependent reactivity of $MoS_2$ and the exceptional chemical inertness of monolayer $MoS_2$ during the nitridation reaction.

Given the stark contrast in electrical properties between semiconducting $MoS_2$ and highly conductive MoN, conductivity maps obtained from MIM measurements (see Methods and Figure S6), provide a powerful tool to monitor the formation and nanoscale spatial evolution of MoN throughout the conversion reaction. As shown in Figure 2c, the pristine $MoS_2$ flake exhibits uniformly low conductivity across all regions, from 1L to 4L, prior to conversion, consistent with the intrinsic semiconducting nature of $MoS_2$.[43,50] At 6 min, a pronounced increase of conductance emerges, indicating the formation of metallic MoN in the corresponding regions.[35,36] The signal is most prominent in the 4L region and along the boundary between the 4L and the 3L domains, suggesting a higher reactivity of these thicker regions compared to the 2L and 1L counterparts. As the reaction progresses, the high-conductance areas in the 4L and 3L regions expand and reach saturation at 13 min, marking their complete conversion to MoN, in agreement with the Raman results. A noticeable conductance increase is observed in the 2L region at 14 min with the contrast continuing to intensify until 18 min, corresponding to the partial and then full conversion of bilayer $MoS_2$. Interestingly, no obvious MIM signal change is detected in the 1L region throughout the conversion, including at 42 min when Raman measurements confirm its full conversion. Additionally, MoN domains derived from thicker $MoS_2$ layers exhibit substantially higher electrical conductance than those converted from thinner layers. This observation implies a layer-dependent variation in the structural quality and morphology of the resulting MoN, highlighting the need for structural and morphological characterizations at nanometer scale to uncover the underlying mechanisms.

To understand the conductivity evolution and elucidate the mechanism underlying the layer-dependent reactivity, we perform TEM characterizations on a fully converted step-shaped $MoS_2$ flake containing regions from 1L to 4L (Figure S7). SAED patterns acquired at various thicknesses uniformly exhibit a single set of hexagonal diffraction spots corresponding to the (100) planes of δ-MoN (Figure 3a and Figure S8), all aligned in the same crystallographic orientation. It should be noted that the emerging of δ-MoN phase is the result of higher conversion temperature (i.e. 680 °C) used to fully convert the $MoS_2$ samples at a higher reaction rate, in contrast to the samples in Figure 1 and 2 where a lower temperature was used for a better control of the reaction rate. This result implies the formation of highly orientated MoN domains with uniform lattice structure and crystallographic alignment across all precursor thicknesses, consistent with the epitaxial atomic substitution mechanism we proposed in previous work.[37] However, high-resolution transmission electron microscopy (HRTEM) images reveal substantial differences in the morphology of the MoN products derived from different initial layer numbers of $MoS_2$ (Figure 3b, Figure S10 to S15). The MoN domains can be readily distinguished from the underlying amorphous SiN membrane due to their higher contrast. In the 1L and 2L regions, the δ-MoN forms discontinuous networks composed of nanoscale crystallites interspersed with voids, resulting in a nanonetwork morphology. In the 3L region, the void density decreases significantly, and the δ-MoN adopts a more continuous and uniform structure. This trend continues in the 4L region, where a nearly void-free, continuous δ-MoN film is observed. This evolution from discontinuous to continuous morphology with increasing precursors thickness provides a structural basis to understand the layer-dependent MIM signals discussed above. The 1L region is highly fragmented with small grains disconnected from each other, leading to very weak signals in MIM, which measures the overall sheet conductance at the 50 ~ 100 nm length scale. While the film is still fractured in the 2L region, the underlying connectivity is much improved, resulting in significantly stronger MIM signals. Starting from 3L and above, the δ-MoN crystals become more continuous and better connected, which explains the much higher sheet conductance in the MIM data.

Importantly, these morphology differences also provide insights into the mechanistic origin of the layer-dependent reactivity. Previous studies have established that the thermodynamically stable δ-MoN phase possesses a unit cell comprising two hexagonal close-packed Mo atomic layers, with nitrogen atoms arranged in an ABAB stacking sequence at the interstitial sites.[51–57] The formation of this structure from a monolayer

MoS$_2$ precursor requires significant atomic rearrangement, including the vertical migration of Mo atoms to form a multilayered configuration along the c-axis. As a result of the atomic rearrangement, MoN crystals converted from monolayer MoS$_2$ precursor are discontinuous, and thus less electrically conductive. Moreover, since a single-unit-cell MoN structure is less stable than those containing multiple Mo atomic layers, the 2L and 3L regions are expected to undergo a certain degree of vertical Mo atom migration to form a more thermodynamically favorable MoN configuration, resulting in voids in the products. Notably, this rearrangement introduces an energetic barrier that renders the transformation less kinetically favorable, particularly in ultrathin regions with limited Mo atoms available in the out-of-plane direction. As a result, thinner MoS$_2$ films, especially monolayers, exhibit reduced reactivity and slower conversion kinetics—consistent with the delayed transformation. Structural reordering of MoS$_2$ layers involving Mo migration and agglomeration has been previously observed under external energy input, such as joule heating[58] and low-energy ion irradiation[59], offering further rational for our proposed mechanism. Figure 3c provides a schematic illustration of the conversion process in 1L and 4L MoS$_2$, highlighting the morphology difference that arises from their distinct structural transformation pathways. In 4L MoS$_2$, nearly no vertical rearrangement of Mo atoms is required to form an energetically favorable MoN structure, thereby preserving the structural continuity of the product. In contrast, in 1L MoS$_2$, substantial vertical rearrangement of Mo atoms is necessary to attain a thermodynamically stable MoN phase, leading to the formation of intergranular voids between MoN nanocrystals. In MoS$_2$ layers of specific thickness, the structural transformation pathway is governed by a trade-off between the thermodynamic stability of the resulting MoN phase and the energetic cost associated with vertical rearrangement of Mo atoms.

To further support our mechanistic interpretation, we quantitatively analyze the lateral coverage of the resulting MoN films based on TEM images (Figure 3d). The conversion process induces a reduction in lateral film coverage due to both vertical Mo atom rearrangement and lattice mismatch between MoS$_2$ and MoN. In the absence of vertical Mo atom rearrangement, the lattice constant difference between MoS$_2$ (i.e. 3.19 Å) and MoN (i.e. 2.89 Å) leads to a projected coverage of $2.89^2 / 3.19^2 = 82.1\%$, corresponding to an 'N-to-N' conversion scenario in which the number of Mo atomic layers is preserved.

If half of the Mo atoms migrate out-of-plane to form an additional Mo layer, the expected lateral coverage decreases to 0.5 × 2.892 / 3.192 = 41.0%, corresponding to an 'N-to-2N' conversion.

The measured MoN coverages for regions converted from 1L and 2L $MoS_2$ are 44.48% and 46.94%, respectively, which are in close agreement with the 'N-to-2N' model. This suggests that monolayer $MoS_2$ is predominantly converted into a MoN structure containing two Mo atom layers, while bilayer $MoS_2$ is converted into a MoN structure containing four Mo atom layers. For the 3L region, the measured MoN coverage is 72.25%, which lies between the values predicted in the 'N-to-N' (82.1%) and '3L-to-4L' (~62%) models. We believe in this case, approximately half of the Mo atoms in 3L $MoS_2$ undergo vertical rearrangement to form a four-Mo-layer MoN, while the other half retain their original plane following the 'N-to-N' model. Our previous work has shown that the MoN structure contains four Mo atom layers exhibits a higher thermodynamic stability compared to three-layer variants,[38] which further support this "3L-to-4L" transformation preference in the 3L case. We also noticed that the measured MoN coverage for 4L region is higher than the predicted values from both the "N-to-N" and "N-to-2N" models. This discrepancy is likely due to residual strain in the MoN film that is not fully released during the conversion process. The model predictions assume complete relaxation of the strain arising from the bond length mismatch between Mo–S and Mo–N. In practice, however, the presence of a supporting substrate makes full strain release difficult to achieve. Moreover, compared with the more interspersed MoN domains formed from thinner $MoS_2$ layers, the more uniform and continuous MoN film derived from 4L $MoS_2$ presents greater challenges for strain relaxation.

In addition to the energy barrier associated with Mo atom rearrangement for the formation of stable MoN phases, another factor that may contribute to the observed layer-dependent chemical reactivity is the adsorption behavior of $NH_3$ on $MoS_2$. To investigate this, we perform density functional theory (DFT) calculations to determine the dependence of $NH_3$ binding on $MoS_2$ layer number. Specifically, we computed the binding energies of a single $NH_3$ molecule on $MoS_2$ layers ranging from 1L to 4L. Given that $MoS_2$ used in our experiments is likely to contain native defects—particularly sulfur vacancies, which are predicted to be the most stable intrinsic defects in MoS$_2$[60] we investigate NH$_3$ adsorption on both pristine and defective surfaces. Four types of defects are considered: 1) a single sulfur vacancy (V$_S$); 2) a defect cluster with three sulfur vacancies (V$_{S3}$); 3) a defect complex consisting of a single Mo vacancy with three neighboring sulfur atoms removed (V$_{MoS3}$); and 4) a larger defect complex involving the removal of three sulfur atoms and one Mo atom from a hexagonal ring (V$_{MoS3, H}$). Structural models of these defects are provided in Figure S16 of the SI. Although the latter two defect complexes are less energetically favorable (see Table S1 in the SI for calculated defect formation energies), they are included to represent potential reactive sites that may exist at step edges or grain boundaries in the experimental samples, or that may form during high-temperature annealing.

Figure 4a presents the relaxed structures of NH$_3$ adsorbed on monolayer MoS$_2$ for both pristine and the four types of defective surfaces. To determine the most favorable adsorption site on pristine MoS$_2$, we consider three possible binding sites: the "hollow" site at the center of a hexagonal ring (labeled "H"), and two on-top sites above Mo (labeled T$_m$) and S (labeled T$_s$) atoms. Additionally, two molecular orientations were evaluated— NH$_3$ pointing "up" (hydrogens away from the surface) and "down" (hydrogens facing the surface). We find that the on-top site with the NH$_3$ molecule in the up orientation corresponds to the minimum energy configuration, with the hollow-up configuration being slightly less favorable by ~15 meV. In the absence of defects, NH$_3$ is only weakly physisorbed on the MoS$_2$ surface, with binding energies ranging from 20 meV to 36 meV. For the defective structures, we focused on NH$_3$ adsorption at the defect site with the molecule oriented down toward the surface. Except for the V$_{MoS3}$ site, NH$_3$ exhibits strong chemsorption at the defect sites, with binding energies ranging from 0.9 eV (V$_{S3}$) to 1.4 eV (V$_S$). In the V$_S$ structure, NH$_3$ undergoes dissociation upon absorption, forming an NH fragment that bonds to the vacancy site through a Mo-N bond, while the remaining two hydrogen atoms remain unbound above the surface. In the V$_{S3}$ case, NH$_3$ dissociates into an NH$_2$ species bonded to the vacancy site through a Mo-N bond and a hydrogen atom bound to an adjacent S atom. These results are consistent with previous experimental observations indicating that defect sites—particularly sulfur vacancies—are required to facilitate NH$_3$ chemisorption,[37] indicating that desulfurization is a prerequisite for MoN synthesis from MoS$_2$ precursors. Figure 4b plots the computed layer dependence of the binding energies for all structures considered. The result shows negligible variation in adsorption energy across different MoS$_2$ thicknesses, regardless of defect type. The pristine MoS$_2$ and V$_{MoS3}$ across all layers exhibit weak physisorption with binding energies of 20–36 meV and ~130 meV, respectively. In contrast, the V$_s$ structure binds NH$_3$ most strongly across all layers (−1.40 eV), followed by V$_{MoS3, H}$ (−1.13 eV) and V$_{S3}$ (−0.90 eV to −1.08 eV). These results clearly indicate that the layer-dependent chemical reactivity observed experimentally cannot be attributed to differences in NH$_3$ binding energy. Instead, it is more likely governed by the extent of rearrangement of Mo atoms necessary to form the thermodynamically stable MoN phases, as discussed in the preceding sections.

**Conclusion**

In conclusion, we have systematically investigated the layer-dependent chemical reactivity of MoS$_2$ in a nitridation reaction, revealing that the energy cost of Mo atom rearrangement governs the observed trend. Monolayer MoS$_2$ exhibits striking inertness due to high kinetic barriers from significant out-of-plane Mo migration during phase transformation. This leads to distinct morphologies: monolayer and bilayer MoS$_2$ form discontinuous MoN nanonetworks with many voids, whereas thicker layers yield more continuous and uniform MoN films. The effect also strongly impacts the local conductance of the resulting thin films. DFT calculations reveal that the transformation is driven by desulfurization and strong Mo–N bond formation, while ruling out NH$_3$ binding as a factor in the layer-dependent reactivity. This work uncovers the atomic-scale mechanisms controlling chemical reactivity and morphology, providing a foundation for tailoring 2D material chemistry in future applications.

**Supporting Information**

Details for the sample preparation, Raman and PL spectra of MoS$_2$ with different number of layers, XPS characterizations, TEM and MIM characterizations, MoN coverage calculations, NH$_3$ absorption energy calculations, additional optical images, statistical analysis on Raman maps, MIM images and TEM images.

**Author Information**

**Corresponding Author**

Xi Ling - Department of Chemistry, Division of Materials Science and Engineering, and The Photonics Center, Boston University, Boston, Massachusetts 02215, United States

Email: xiling@bu.edu

**Authors**

Zifan Wang - Department of Chemistry, Boston University, Boston, Massachusetts 02215, United States

Jiaxuan Wen - Department of Physics, The University of Texas at Austin, Austin, Texas 78712, United States

Tina Mihm - Department of Electrical and Computer Engineering, Boston University, Boston, Massachusetts 02215, United States

Shaopeng Feng - Department of Physics, The University of Texas at Austin, Austin, Texas 78712, United States

Kelvin Huang - Department of Electrical and Computer Engineering, Boston University, Boston, Massachusetts 02215, United States

Jing Tang - Division of Materials Science and Engineering, Boston University, Boston, Massachusetts 02215, United States

Tianshu Li - Division of Materials Science and Engineering, Boston University, Boston, Massachusetts 02215, United States

Liangbo Liang - Center for Nanophase Materials Sciences, Oak Ridge National Laboratory, Oak Ridge, Tennessee 37831, United States.

Sahar Sharifzadeh - Department of Electrical and Computer Engineering, Boston University, Boston, Massachusetts 02215, United States

Keji Lai - Department of Physics, The University of Texas at Austin, Austin, Texas 78712, United States

**Notes**

The authors declare no competing financial interests.

**Acknowledgements**

Research is primarily supported by the U.S. Department of Energy (DOE), Office of Science, Basic Energy Sciences (BES), under Award DE-SC0021064. Work by X.L. was also supported by the National Science Foundation (NSF) under Grant Nos. (1945364) and (2111160). The microwave microscopy work (J.W., S.F., K.L.) was supported by the Welch Foundation (Grant No. F-1814). K.L. was also supported by NSF under Grant DMR-2426989. Computational studies by T.M, K.H, and S.S. were supported by DOE BES under Award number DE-SC0023402. L.L. acknowledges work at the Center for Nanophase Materials Sciences, which is a U.S. Department of Energy Office of Science User Facility. This work was performed in part at the Harvard University Center for Nanoscale Systems (CNS), a member of the National Nanotechnology Coordinated Infrastructure Network (NNCI), which is supported by the National Science Foundation under NSF award no. ECCS-2025158. We would like to acknowledge computational resources from the National Energy Research Scientific Computing Center (NERSC), a DOE Office of Science User Facility supported by the Office of Science of the U.S. Department of Energy under Contract No. DE-AC02-05CH11231, and Boston University's Research Computing Services.

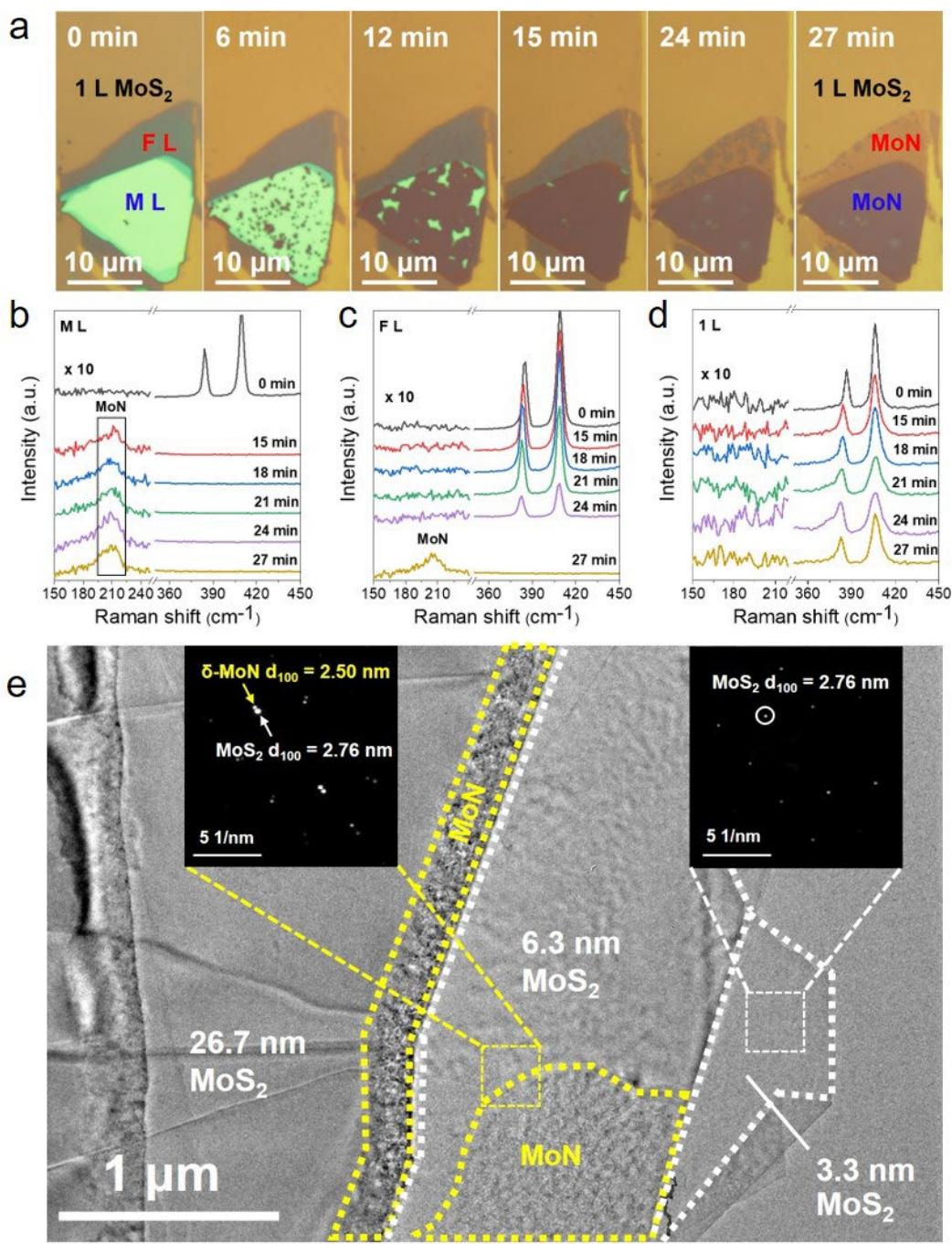

**Figure 1. Demonstration of the chemical reactivity differences among 1L, FL and ML MoS$_2$.** (a) Optical images of a step-shaped MoS$_2$ flake at various stages of the nitridation atomic substitution reaction. (b-d) Corresponding Raman spectra measured in the ML (b), FL (c) and 1L region (d) of the MoS$_2$ flakes in (a). (e) TEM image of a partially converted step-shaped MoS$_2$. Thickness step edges between regions of different layer numbers are indicated by white dashed lines, while areas converted to MoN are outlined in yellow. Insets are SAED patterns acquired from the marked locations.

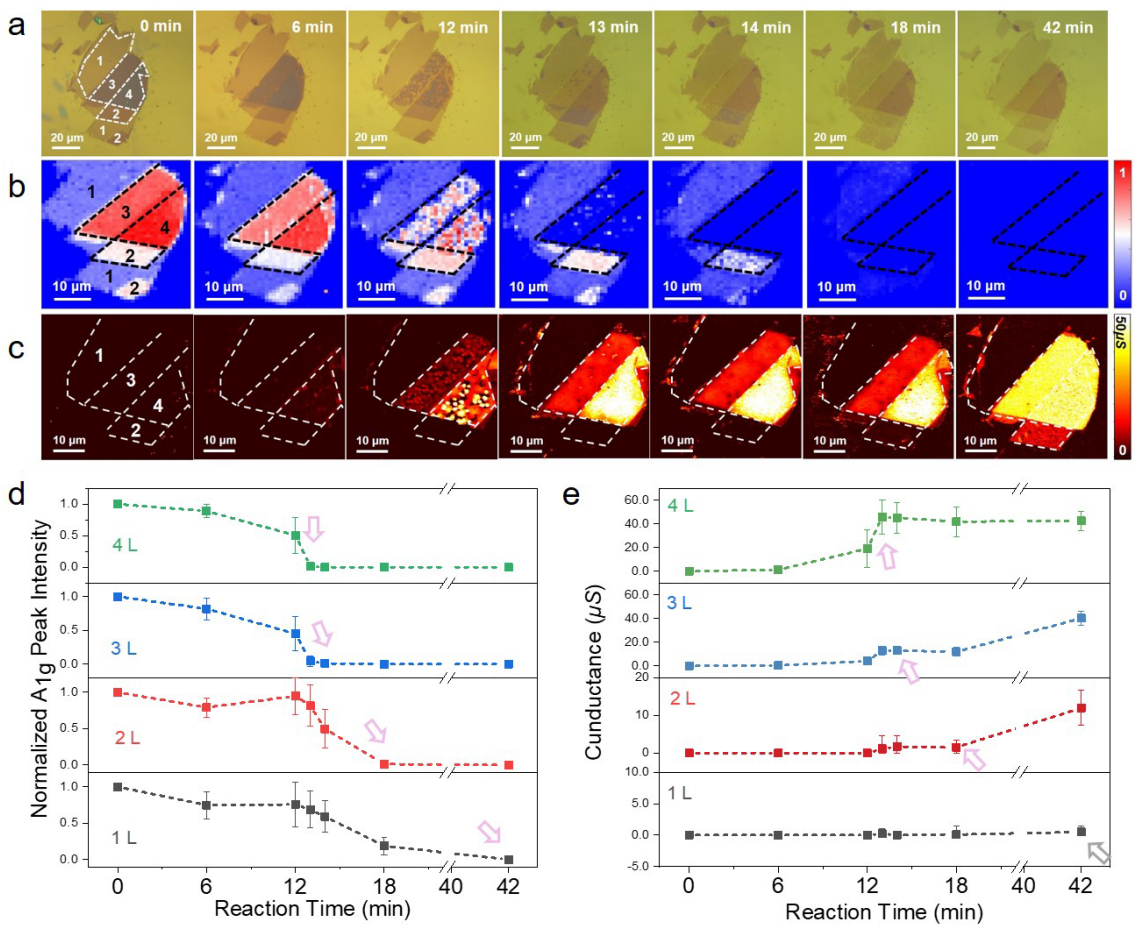

**Figure 2. Investigation of the layered-dependent chemical reactivity on ultrathin MoS$_2$ ranging from 1L to 4L.** (a) Optical images, (b) Raman intensity maps of the A$_{1g}$ mode of MoS$_2$, (c) conductance maps from the MIM measurements of an ultrathin step-shaped MoS$_2$ flake at different conversion stages. The 1L, 2L, 3L and 4L regions are labeled as 1, 2, 3 and 4, respectively. Boundaries between adjacent thickness regions are indicated by white dashed lines. (d) Time-dependent evolution of the average A$_{1g}$ peak intensity extracted from the Raman maps shown in (b). The average A$_{1g}$ intensity is normalized to its initial value prior to the reaction. (e) Time-dependent evolution of the average sheet conductance values extracted from the maps shown in (c). The pink arrows indicate the shift in the duration required to complete the conversion, as inferred from changes in both Raman intensity and electrical conductance. The gray arrow in (e) marks a hypothetical point where conductance increase would occur in the 1L sample.

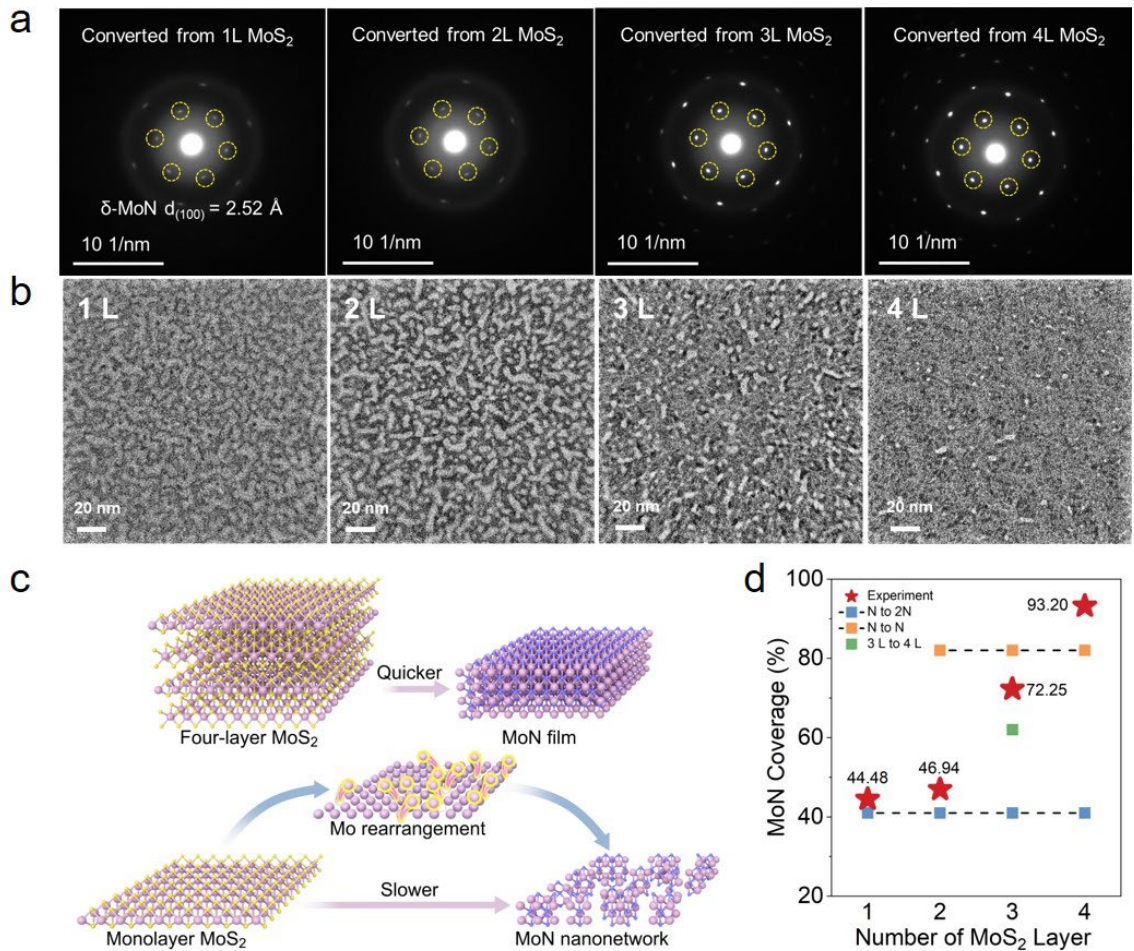

**Figure 3. Structural and morphological analysis of MoN formed from the nitridation of 1L to 4L MoS$_2$, and the proposed structural transformation pathways.** (a) SAED patterns and (b) TEM images of MoN crystals from 1L, 2L, 3L, and 4L MoS$_2$, respectively. (c) Schematic illustration of the distinct structural transformation pathways from 1L and 4L MoS$_2$ to their corresponding MoN structures. (d) MoN film coverages obtained experimentally (star symbols) and theoretically (square symbols) from the conversion of 1L to 4L MoS$_2$.

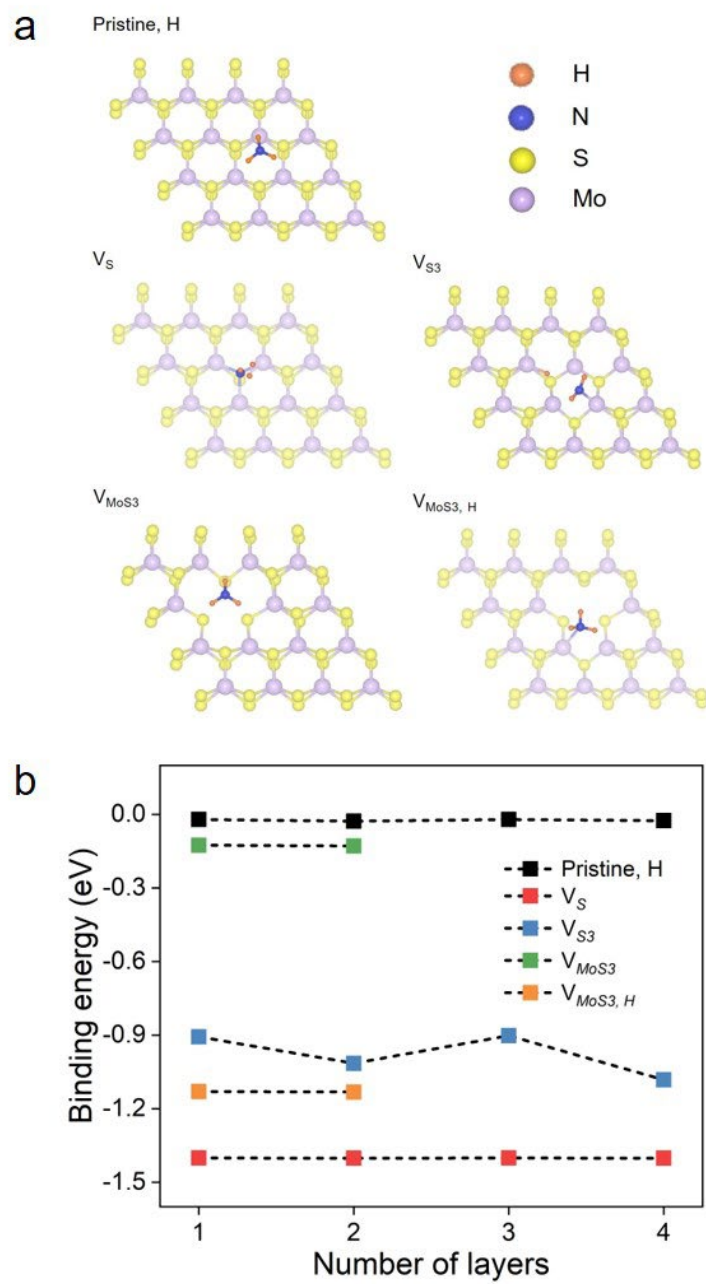

**Figure 4. Structures and binding energies of $NH_3$ adsorbed on pristine and defective $MoS_2$ surfaces with varying layer numbers.** (a) The optimized structure for a $NH_3$ molecule adsorbed on monolayer $MoS_2$. (b) Variation of $NH_3$ binding energy with increasing number of $MoS_2$ layers.

Supplementary Materials for

# Mo Atom Rearrangement Drives Layer-Dependent Reactivity in Two-Dimensional MoS$_2$

Zifan Wang[1], Jiaxuan Wen[2], Tina Mihm[3], Shaopeng Feng[2], Kelvin Huang[3], Jing Tang[4], Tianshu Li[4], Liangbo Liang[6], Sahar Sharifzadeh[3], Keji Lai[2], Xi Ling[1, 4, 5*]

[1] Department of Chemistry, Boston University, Boston, Massachusetts 02215, USA.

[2] Department of Physics, The University of Texas at Austin, Austin, Texas 78712, USA.

[3] Department of Electrical and Computer Engineering, Boston University, Boston, Massachusetts 02215, USA.

[4] Division of Materials Science and Engineering, Boston University, Boston, Massachusetts 02215, USA.

[5] The Photonics Center, Boston University, Boston, Massachusetts 02215, USA.

[6] Center for Nanophase Materials Sciences, Oak Ridge National Laboratory, Oak Ridge, Tennessee 37831, USA.

*To whom the correspondence should be addressed. Email address: xiling@bu.edu

**Methods**

*Nitridation Atomic Substitution Reaction of MoS$_2$*

MoS$_2$ flakes were prepared via mechanical exfoliation of bulk single crystals and transferred onto substrates such as SiO$_2$/Si. The sample were then loaded into a 1-inch quartz tube housed within a tube furnace. For the atomic substitution reaction, the furnace temperature was ramped to 650 °C at a rate of 45 °C/min and held at the target temperature for various durations for the reaction. A mixture of NH$_3$ gas and argon was introduced into the system as the nitrogen source for the atomic substitution. The flow rate of each gas is 50 sccm. After the reaction, the furnace was cooled naturally to room temperature before retrieving the samples for subsequent characterization using various analytical techniques.

*TEM Sample Preparation and Characterization*

MoS$_2$ flakes of interest are transferred from the SiO$_2$/Si substrate onto silicon nitride TEM grids via a dry-transfer method. Specifically, a drop of polystyrene (PS) in toluene (20% wt) is casted and spin coated on the SiO$_2$/Si substrate containing the target MoS$_2$ flake. The substrate is then heated on a hot plate at 85 °C for 4 mins, followed by immersion in deionized (DI) water for 5 mins. Subsequently, the PS film—along with the embedded MoS$_2$ flake—is lifted from the substrate using tweezers and transferred onto a polydimethylsiloxane (PDMS) stamp, with the flake-oriented face-up. The PDMS stamp is then brought into contact with the TEM grid and heated at 95 °C for 10 minutes, during which the PS film and MoS$_2$ flake detach from the PDMS and adhere to the grid. To remove the PS film, the TEM grid is soaked in toluene for 20 minutes. Finally, a 90-minute nitridation reaction at temperature of 680 °C is performed to fully convert the MoS$_2$ flake to MoN. TEM measurements are performed on JEOL ARM 200F STEM in the Center for Nanoscale Systems (CNS) at Harvard University.

*Raman and Photoluminescence (PL) Measurements*

Raman and PL measurements were performed using a Renishaw inVia Raman microscope equipped with a 532 nm laser excitation. Raman mappings were performed over an area of 40 μm × 48 μm at various stages of the conversion process, with a step size of 1 μm for both X-aixs and Y-axis. In order to account for potential variations in signal intensity, Raman measurements were also acquired from a reference $SiO_2$/Si substrate. The intensity of the Si peak at 520 cm$^{-1}$ from the reference substrate is denoted as $I_{Si-Xmin}$, where X presents the reaction time in min. Calibration and normalization were then applied to each point in all Raman maps using the following equation:

$$I_{C-Xmin} = I \times (I_{Si-0min} / I_{Si-Xmin})$$

$$I_{N-Xmin} = I_{C-Xmin} / I_{max-0min}$$

Here, $I$ is the original intensity measured at X min, $I_{C-Xmin}$ is the calibrated intensity using the reference $SiO_2$/Si substrate, $I_{max-0min}$ is the highest intensity value within the map acquired at 0 min, and $I_{N-Xmin}$ is the normalized intensity to the calibrated intensity value at 0 min. The average and standard deviation of $I_{N-Xmin}$ in each of the four thickness regions (1L to 4L) were calculated and plotted as a function of the reaction time. The results are presented in Figure S4.

*Statistical Analysis of Raman Map*

The relative intensity of each point ($I_{R-Xmin}$) in the Raman maps was calculated using the formula shown below:

$$I_{R-Xmin} = I_{N-Xmin} / I_{N-0min}$$

The average and standard deviation of $I_{R-Xmin}$ in each of the four thickness regions were calculated and plotted as a function of the reaction time. The result is presented in Figure 2d.

*Microwave Impedance Microscopy (MIM) Measurements and Finite-Element Analysis (FEA)*

The MIM experiments in this work were performed on an AFM platform (ParkAFM XE-70) using tapping-mode AFM cantilevers with an overall Au coating (160AC-GG) from OPUS.[1] Before the measurement, the mixer phase was adjusted such that the signal in the MIM-Re channel is zero on the insulating $SiO_2$ substrate. The measured MIM-Re and MIM-Im images at different stages of the conversion process in Fig. 2 in the main text are shown in Fig.S5(a) and (b), respectively. The commercial software COMSOL 5.4 was used to perform finite element analysis (FEA) to simulate the real and imaginary parts of the admittance for the specific tip−sample geometry in Fig. S5(c). In the experiment, the cantilever tip oscillates sinusoidally with an amplitude of 50 nm and a frequency of 250.7 kHz. As shown in Fig.S5(d), by mapping the tip oscillation onto the MIM approach curve, we can establish the relationship between the actual oscillating signal and time. Applying a Fast Fourier Transform (FFT) to this time-domain signal, we obtain the simulated AC MIM response as a function of sheet conductance in Fig.S5(d). In this study, the simulated ratio of MIM-Re to MIM-Im was calculated and compared with experimental data to form the local sheet conductance images of the $MoS_2$/MoN sample, as shown in Fig.2c in the main text.

*Estimation of MoN Film Coverage from TEM Analysis*
<u>Image Processing:</u>
To improve image contrast and correct uneven illumination, Contrast Limited Adaptive Histogram Equalization (CLAHE)[2] was applied in the CIELAB color space. The original image was first converted from BGR to LAB color space to isolate the luminance channel (L) from the chromatic channels (A and B). CLAHE was applied exclusively to the L channel using a clip limit of 10 and a tile grid size of 10 × 10, allowing for localized contrast enhancement while minimizing noise amplification. The enhanced L channel was then recombined with the original A and B channels to reconstruct the LAB image, which was subsequently converted back to the BGR color space. This method preserves chromatic integrity while effectively correcting spatial variations in lighting.

*Coverage Estimation:*

To quantitatively estimate the areal coverage of MoN film, the contrast-enhanced image was first converted to grayscale and binarized using a global intensity thresholding method. An empirically determined threshold was applied via the inverse binary thresholding function to distinguish the darker material region (e.g. MoN) from the lighter background (e.g. blank TEM grid region). In the resulting binary image, pixels corresponding to the material were assigned a value of 255 (white), and background pixels were set to 0 (black). The material coverage percentage was then computed as the ratio of material pixels ($N_{material}$) to the total number of pixels in the image ($N_{total}$), expressed as:

$$Coverage\ Rate = \left(\frac{N_{material}}{N_{total}}\right) * 100\%$$

This approach provides a robust, reproducible means to quantify the surface coverage of materials following contrast enhancement and noise suppression.

*Theoretical and Computational Calculations*

First-principles density functional theory (DFT) calculations were performed with a planewave basis as implemented in the Vienna Ab-initio Simulation Package (VASP).[3–7] All calculations were run with VASP 6.4.2 using the Perdew–Burke–Ernzerhof (PBE)[8] generalized gradient approximation functional to treat the exchange-correlation energy. We employed frozen core projector-augmented wave (PAW) potentials to describe the nuclei and core with default pseudopotentials for VASP.[9,10] For the unit cell, we utilized a 6 × 6 × 1 k-point mesh with a planewave cutoff of 388 eV, which converged the total energy below 1 meV/atom. Geometry optimizations were done using the RMM-DIIS[3,4,11] algorithm, where all atoms were allowed to optimize until all forces were converged to less than 0.01 eV/ Å.    The lattice vectors kept to that of the unit cell and were not re-optimized for the defective supercells. The lattice vectors of our unit cell were $(3.181689, 3.181689, 14.838772)$ Å, with cell angles of $(90°, 90°, 120°)$.

The NH$_3$ molecule was adsorbed on the surface of a supercell, with the size of supercell increased as the size of defect was increased. All systems had a 10 Å vacuum layer in the c direction to avoid inter-layer interactions. For the sulfur vacancies (V$_S$ and V$_{S3}$), a $6 \times 6 \times n$ supercell was utilized in order to reduce interactions between defect sites, whereas for the V$_{MoS3}$ structure an $8 \times 8 \times n$ supercell was found to be necessary based on strain analysis at the edge of the supercell. Due to the larger computational cost for V$_{MoS3}$, we only considered this configuration with 1L and 2L MoS$_2$. Supercell calculations were performed with a $1 \times 1 \times 1$ k-point mesh for all defect configurations except for V$_{S3}$ 1L and 2L configurations for which we utilized a $2 \times 2 \times 1$ k-point mesh because this k-point mesh was found to be necessary to obtain the lowest energy atomic structure.

The binding energy for an NH$_3$ molecule adsorbed onto the MoS$_2$ surface was calculated as,

$$E_{bind} = E_{MoS2-NH3} - E_{MoS2} - E_{NH3},$$

where $E_{MoS2-NH3}$ is the energy of the MoS$_2$ with the NH$_3$ on the surface, $E_{MoS2}$ is the energy of the MoS$_2$ surface, and $E_{NH3}$ is the energy of the NH$_3$ molecule. We note that by this definition a negative E$_{bind}$ indicates binding and positive indicates an unbound molecules.

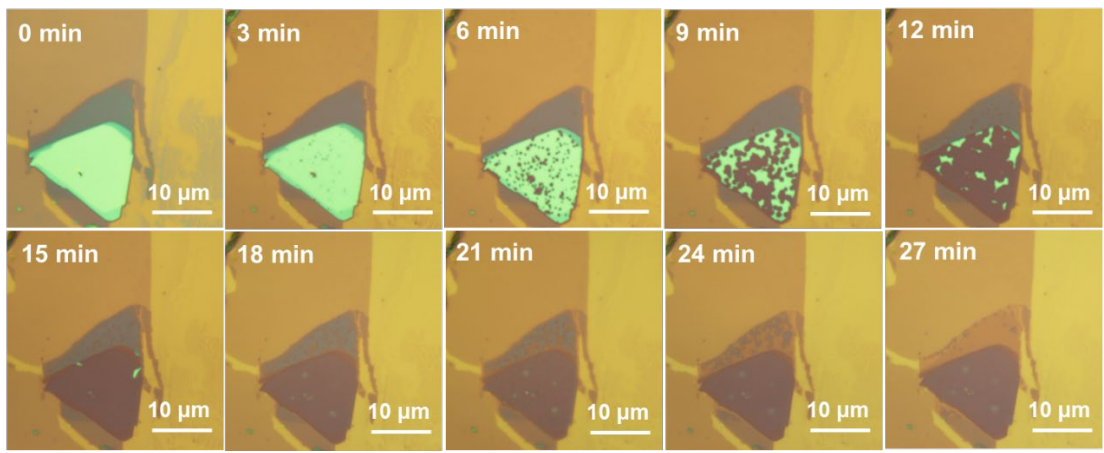

**Figure S1.** Optical images of the step-shaped MoS₂ at various stages of the nitridation atomic substitution reaction.

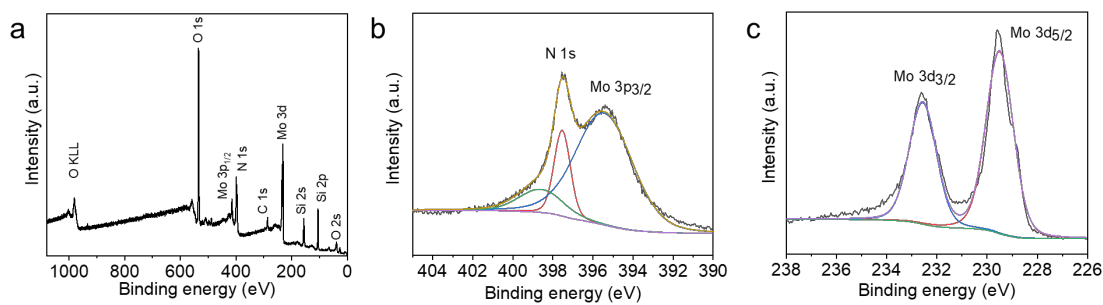

**Figure S2.** (a) XPS survey spectrum of the fully converted sample. (b-c) Fine scan of the XPS spectrum in the (b) Mo3p and N 1s region and (c) Mo 3d region.

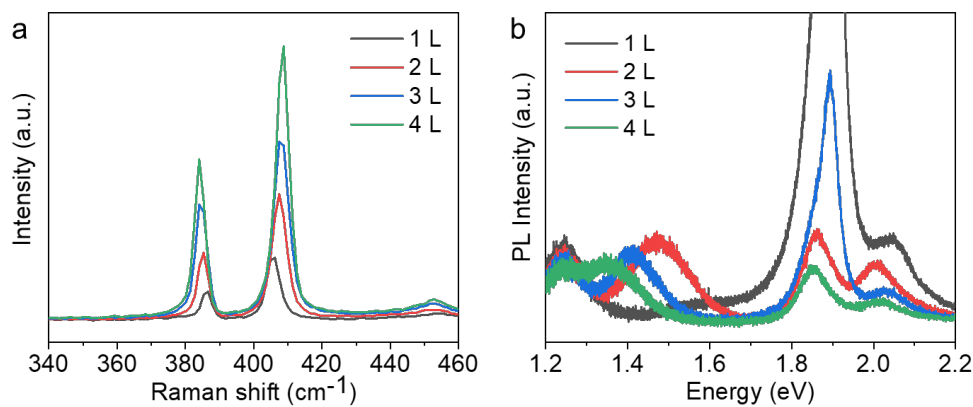

**Figure S3.** Raman (a) and PL (b) spectra measured at 1L, 2L, 3L and 4L regions of the ultrathin step-shaped MoS₂ sample.

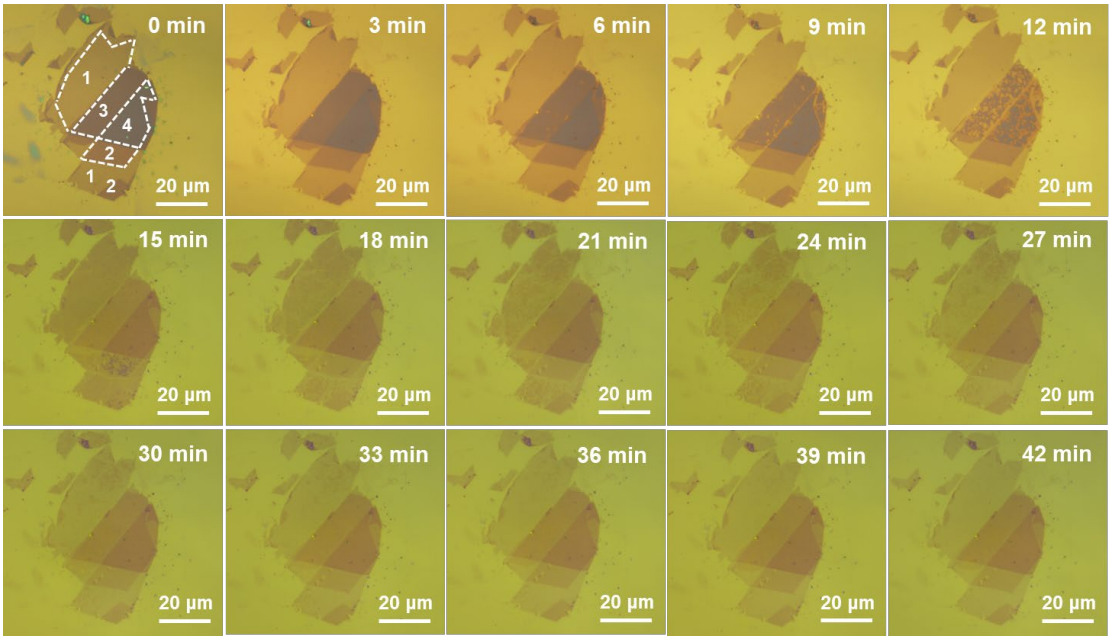

**Figure S4.** Optical images of the ultrathin step-shaped MoS$_2$ sample at various stages of the nitridation atomic substitution reaction.

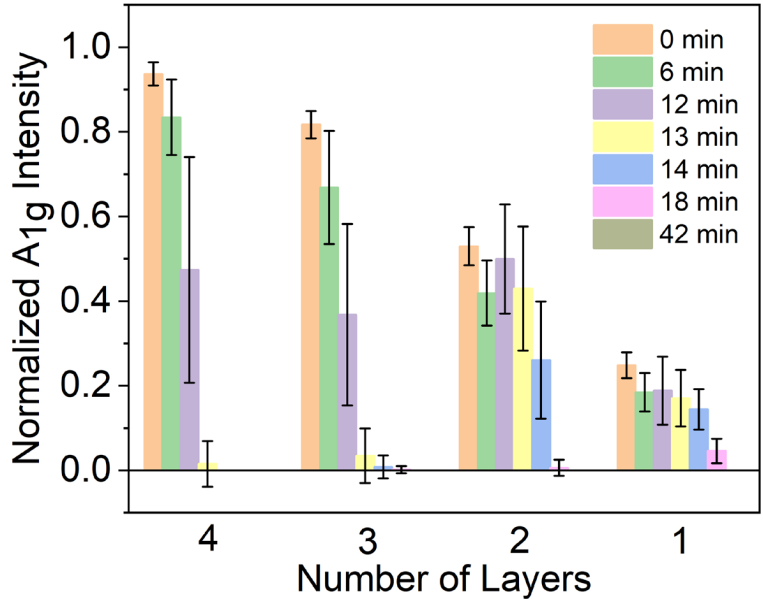

**Figure S5.** Histogram of the normalized integrated A$_{1g}$ intensity of each thickness region of the ultrathin step-shaped MoS$_2$ sample at various stages of the reaction.

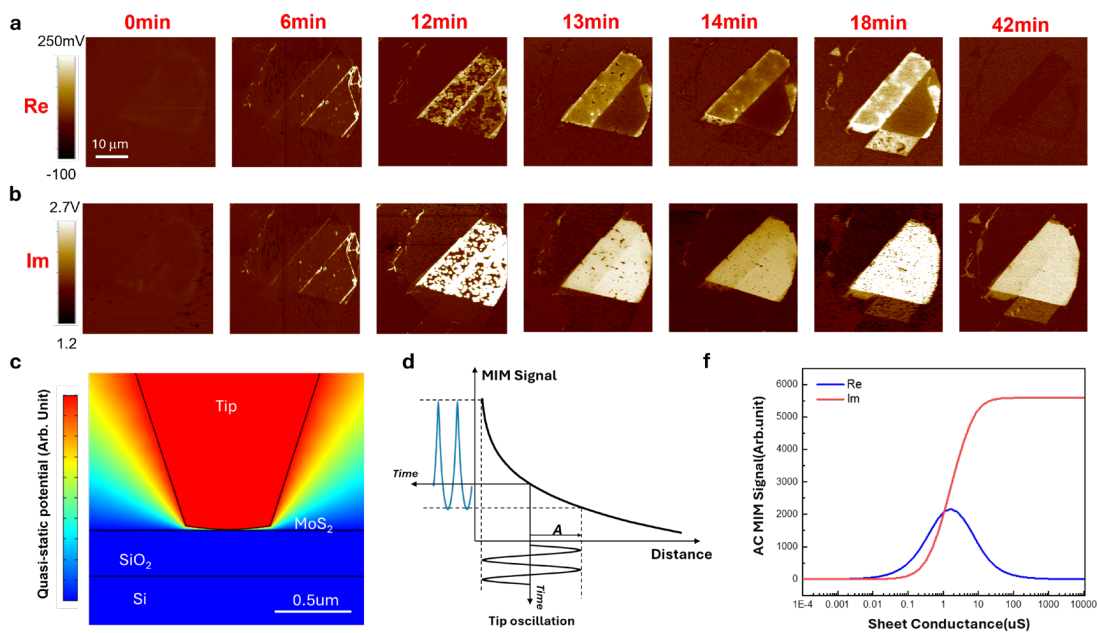

**Figure S6**. MIM characterization and analysis of the MoS$_2$ to MoN conversion process with different number of layers. (a) MIM-Re and (b) MIM-Im images of the sample at different stages of the conversion process. (c) Geometry and quasi-static potential distribution in the COMSOL simulation. (d) Simulated DC MIM signal as a function of tip-sample distance, from which the demodulated AC signals can be extracted. (d) Simulated AC MIM signals as a function of 2D sheet conductance.

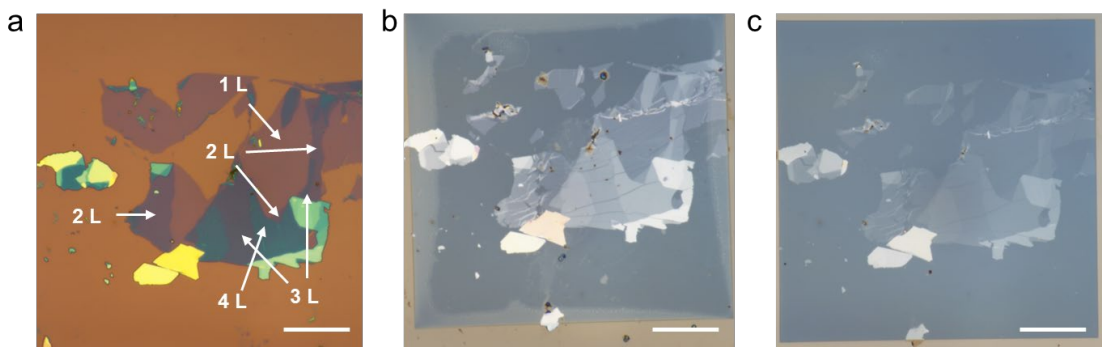

**Figure S7.** Optical images of the MoS$_2$ sample for TEM characterization. (a) On a SiO$_2$/Si substrate; (b) After being transferred to a TEM grid; (c) After nitridation reaction for 90 mins. Scale bars are 10 μm.

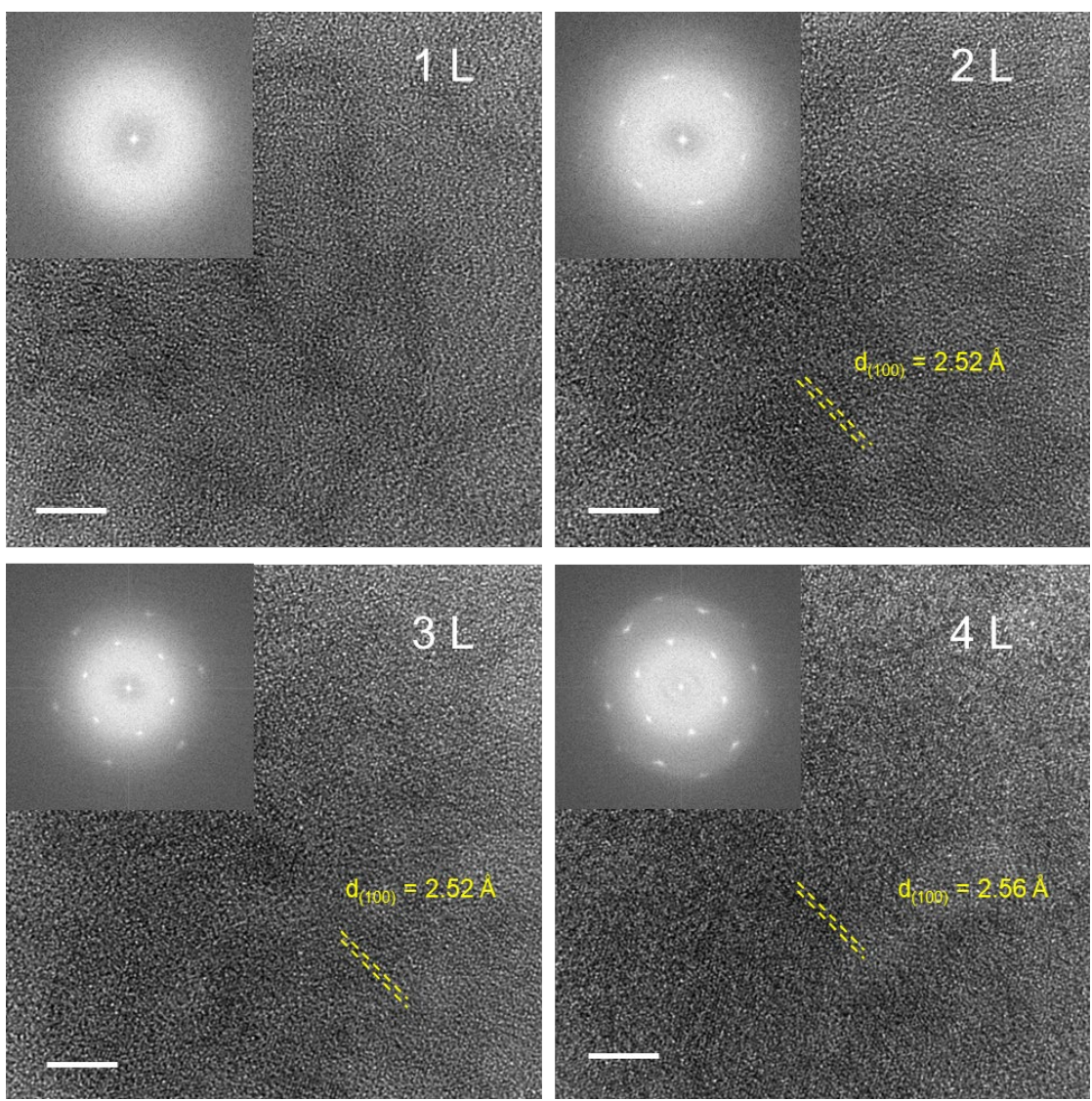

**Figure S8.** HR-TEM images of MoN crystals converted from 1L, 2L, 3L, and 4L MoS$_2$, with the d-spacing of δ-MoN (100) planes labeled. Insets are corresponding fast Fourier transform (FFT) patterns. Scales bars are 5 nm.

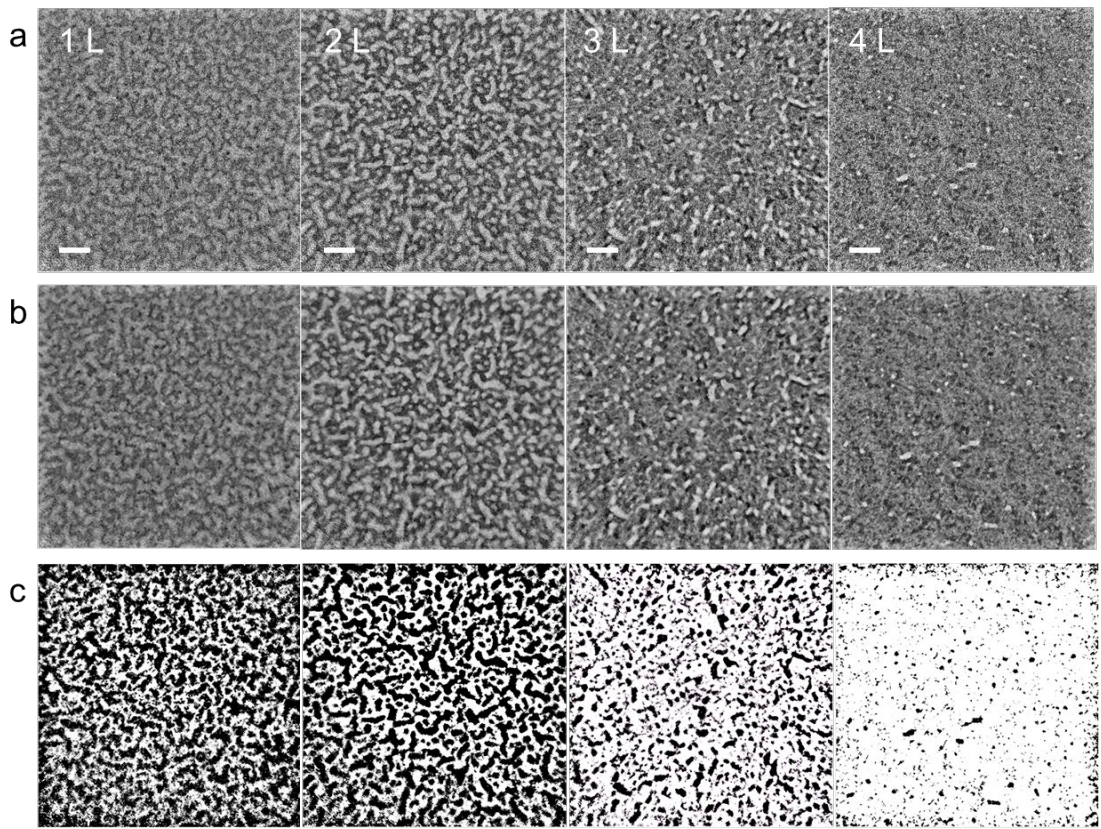

**Figure S9.** HR-TEM images of MoN structures converted from 1L, 2L, 3L, and 4L MoS$_2$. (a) Original TEM images; (b) Image after CLAHE-based contrast and illumination correction; (c) Final processed image used for coverage calculation. Scales bars are 20 nm.

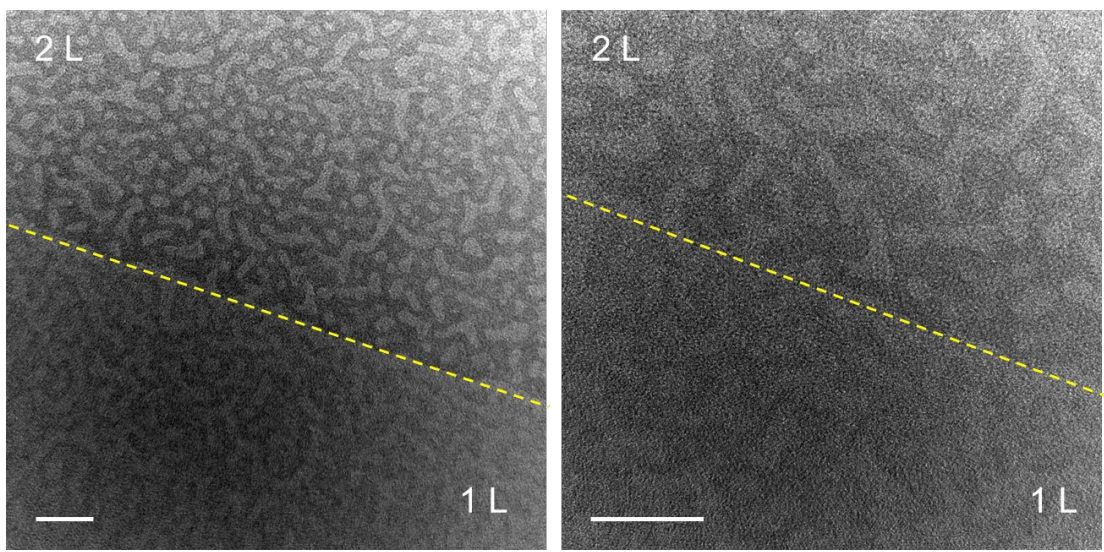

**Figure S10.** TEM images measured at the boundaries between MoN structures converted

from 1L and 2L MoS$_2$. Scale bars are 20 nm.

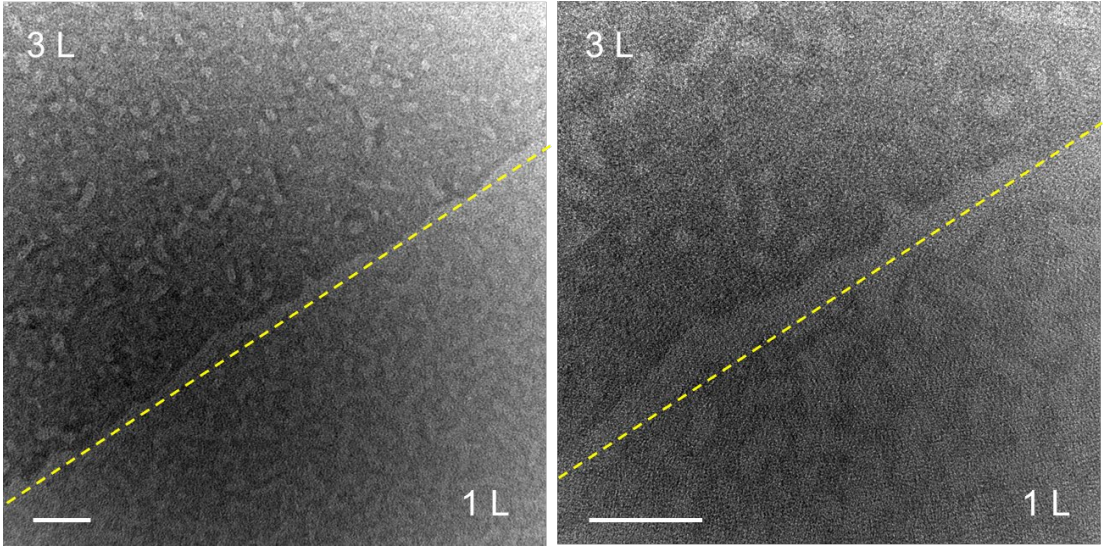

**Figure S11.** TEM images measured at the boundaries between MoN structures converted from 1L and 3L MoS$_2$. Scale bars are 20 nm.

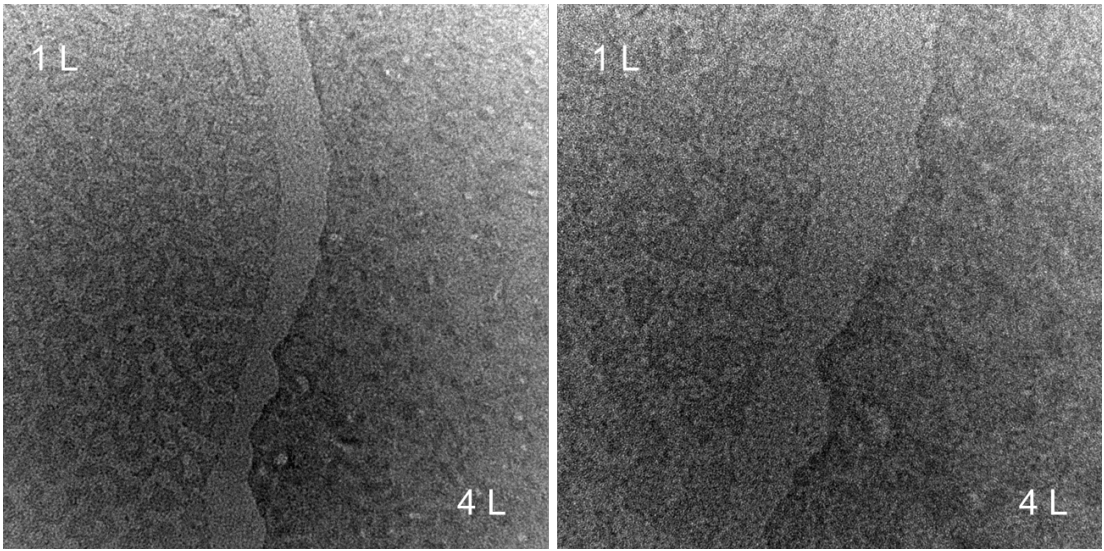

**Figure S12.** TEM images measured at the boundaries between MoN structures converted from 1L and 4L MoS$_2$. Scale bars are 20 nm.

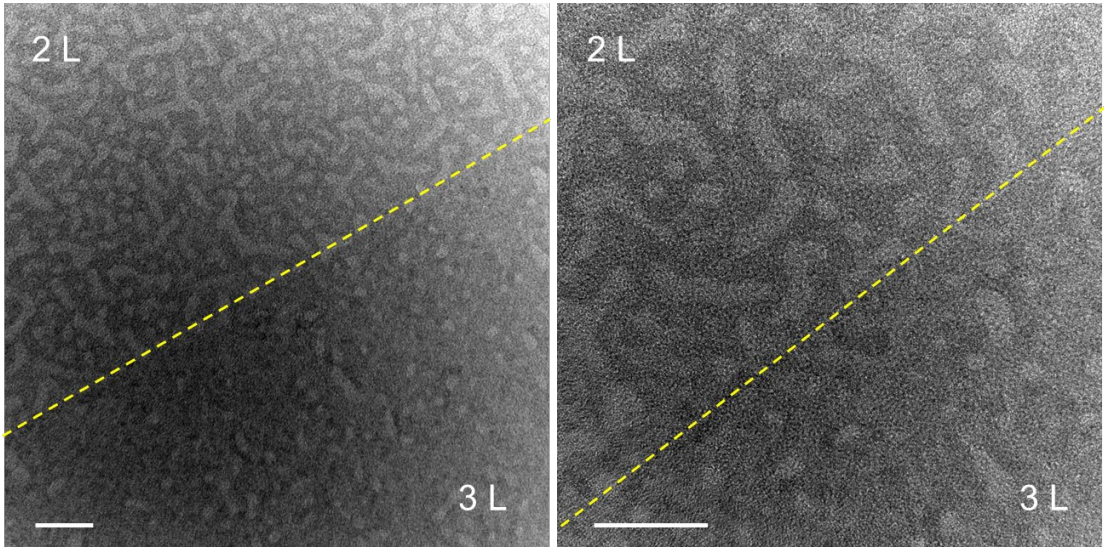

**Figure S13.** TEM images measured at the boundaries between MoN structures converted from 2L and 3L MoS$_2$. Scale bars are 20 nm.

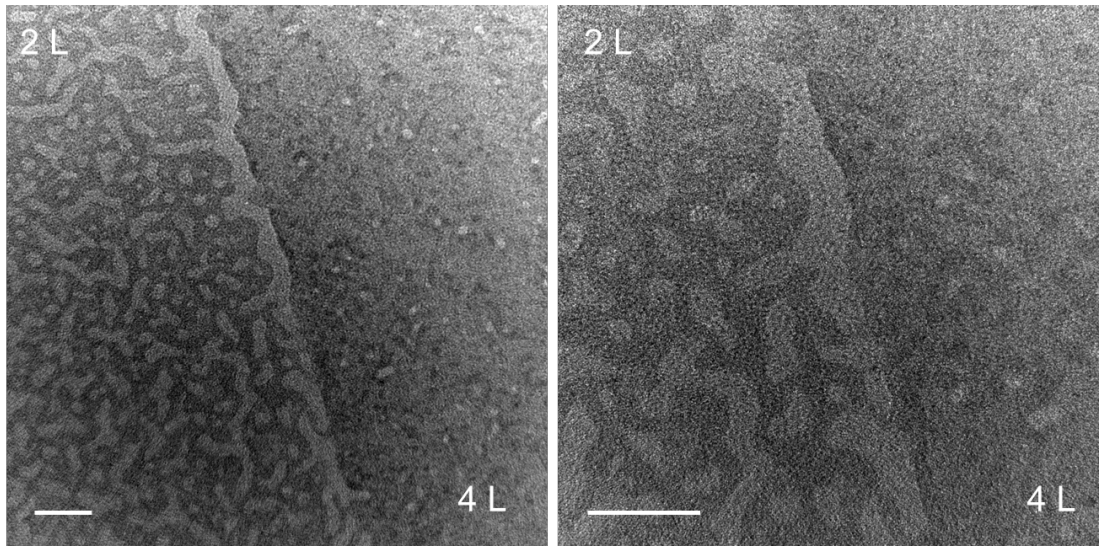

**Figure S14.** TEM images measured at the boundaries between MoN structures converted from 2L and 4L MoS$_2$. Scale bars are 20 nm.

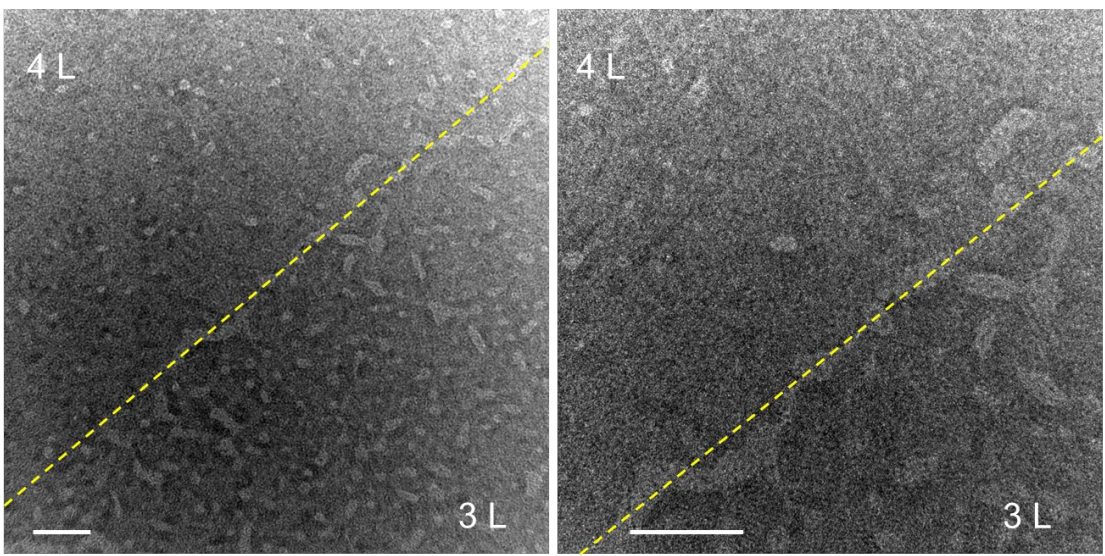

**Figure S15.** TEM images measured at the boundaries between MoN structures converted from 3L and 4L MoS₂. Scale bars are 20 nm.

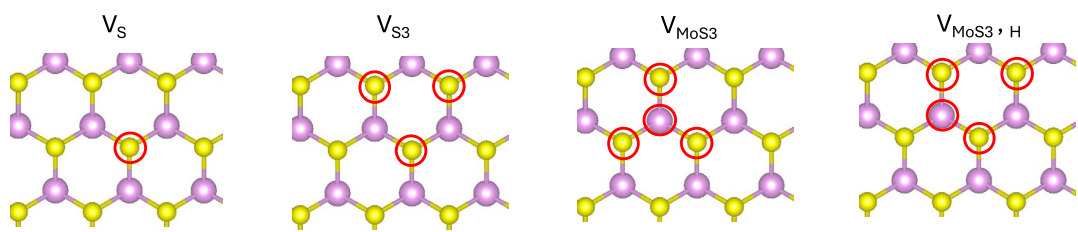

**Figure S16.** Schematics showing the vacancy position for each defect structure. The red circles indicate that atom was removed from the MoS₂ surface to create a vacancy.

Figure S15 shows the four defect complexes studied with DFT. Table S1 reports the formation energies for each of these defects calculated as,

$$E_{form} = E_{MoS2+V} - E_{MoS2} + (N_S \times E_S) + (N_{Mo} \times E_{Mo})$$

where $E_{MoS2+V}$ is the energy of the MoS₂ surface with the vacancy, $E_{MoS2}$ is the energy of the pristine MoS₂ surface, $E_S$ ($E_{Mo}$) is the energy of a single S (Mo) atom, $N_S$ ($N_{Mo}$) is the number of sulfur (molybdenum) atoms removed. As seen in the table, V$_S$ has the lowest formation energy, and is therefore the most likely defect, followed by V$_{S3}$ and the two V$_{MoS3}$ defects. This finding is consistent with previous studies on MoS₂ defects.[12] We note that for the formation energies, a $3 \times 3 \times n$ supercell is sufficiently converged for the single sulfur vacancy, so it is used in place of the larger $6 \times 6 \times n$ supercell used for all binding energy calculations in the work.

**Table S1.** Calculated formation energies for the four vacancy-type point defects considered in this study.

| *Defect* | **Supercell** | **E$_{form}$ (eV)** |
|---|---|---|
| $V_S$ | 331 | 2.6 |
| $V_{S3}$ | 661 | 8.08 |
| $V_{MoS3}$ | 881 | 9.6 |
| $V_{MoS3}$, $H$ | 881 | 11.43 |